\documentclass[twocolumn,showpacs,reprint,amsmath,amssymb,aps,prb]{revtex4-1}
\usepackage{graphicx}
\usepackage{dcolumn}
\usepackage{bm}
\usepackage{natbib}
\usepackage{multirow}
\usepackage{soul}
\usepackage[usenames,dvipsnames]{xcolor}
\usepackage{soul}
\usepackage{floatrow}
\usepackage{amsmath}
\usepackage{amsfonts}
\usepackage{subcaption}
\usepackage[hyperfootnotes=false]{hyperref}
\usepackage{epstopdf}
\hypersetup{
 colorlinks=true,
 citecolor=blue,
 linkcolor=blue,
 urlcolor=Blue}
\begin{document}
\title{Composition and Structure Based GGA Bandgap Prediction Using Machine Learning Approach}

\author{Mukesh K. Choudhary\textit{$^{1,2}$}  Amal Raj V\textit{$^{2}$}, Gowri Sankar S\textit{$^{1,2}$} and P. Ravindran\textit{$^{1,2}$} }
\email{raviphy@cutn.ac.in}

\affiliation{$^{1}$Department of Physics, School of Basics and Applied Sciences, Central University of Tamil Nadu, Thiruvarur, India}
\affiliation{$^{2}$Simulation Center for Atomic and Nanoscale MATerials (SCANMAT) Central University of Tamil Nadu, Thiruvarur, India.}

\begin{abstract}
The focus of this study is to develop  machine learning (ML) regression models to predict the energy bandgap value from chemical compositions and crystal structure with accuracy of those obtained from GGA$-$PBE calculations and validate the prediction through density functional theory (DFT) based band structure calculations.
To achieve this goal, we conducted a comparative analysis of the performance of eight distinct stand-alone ML regression models with huge dataset. Additionally, we developed four ensemble models using the stacking method and seven ensemble models using the bagging method incorporating Ridge regression with cross validation(RidgeCV) and Least Absolute Shrinkage and Selection Operator with cross validation(LassoCV), respectively. All these models were aimed at predicting the GGA$-$PBE bandgap values from chemical compositions and crystal structure with the help of huge number of compounds and their structure from matminer~\cite{ward2018matminer} dataset  matbench$-$mp$-$gap. 
This dataset consists of bandgap values of 106,113 compounds obtained from GGA$-$PBE calculations. For the present analysis the dataset was divided into subsets of increasing sizes: 10000, 25000, 50000 and 100000 entries. All the  NAN values were dropped to clean these dataset resulting 9,223, 23,173, 46,528, and 94,815 entries, respectively, 
 The stand-alone ML regression models such as AdaBoost, Bagging, CatBoost, Light Gradient Boosting Machine(LGBM), Random Forest (RF), decision tree(DT), Gradient Boosting(GB), and eXtreme Gradient Boosting(XGB), were analyzed in terms of their performance for GGA$-$PBE bandgap value prediction across diverse material structures and compositions using above mentioned cleaned datasets with increasing data sizes.
The study utilized matminer and Pymatgen  to featurize the datasets. The permutation of feature importance technique was adopted to identify important features and Pearson correlation method for correlation coefficient matrix.
Among the eight stand-alone models tested, RF model demonstrated the highest performance with an R$^2$ value of 0.943 and RMSE value of 0.504 eV. Next to RF regression, the bagging regression performed better across varying sample sizes if we use streamlined feature selection. 
Importantly, if we employ ensemble models to predict GGA$-$PBE bandgap value, we found that the bagging models demonstrated relatively good performance compared to stand-alone models, achieving best R$^2$ value of 0.948 and RMSE value of 0.479 eV in the test dataset.  
With the best performing trained model we have predicted bandgap value of  new half$-$Heusler compounds with 18 valence electron count and validated the ML prediction using accurate DFT calculations. The calculated bandgap values for these new compounds from DFT calculation are found to be comparable with our best ML regression model predictions. The calculated electronic and phonon band structures show that these new compounds are narrow bandgap semiconductors and are dynamically stable as their all-phonon dispersion curves have positive frequencies.
From this study, we conclude that using the currently trained ensemble method of bagging models one can predict the bandgap values with the accuracy obtained from the GGA$-$PBE method without much computational effort and hence accelerate the discovery of potential optoelectronic materials from wide chemical space.
\end{abstract}
\maketitle
\section{Introduction}
Machine Learning (ML) has emerged as a powerful tool in the field of data analytics, enabling academia and industry professionals to glean insights and make predictions based on complex and varied data sources~\cite{sarker2021machine}. Both academia and industry professionals share a common aim of more accurate and effortless implementation of ML techniques to design advanced functional materials.~\cite{schleder2019dft, lee2021technological,ahmad2022data,guan2020artificial}. In academia, researchers are exploring new ML techniques and algorithms, while in industry, there is a growing demand for professionals with expertise in ML. But both face challenges  as inadequacy of large amounts of high$-$quality data and ensuring ML models are transparent and explainable, particularly in applications such as healthcare where biased or incomplete data can lead to inaccurate predictions and serious aftereffects~\cite{esmaeilzadeh2020use,reddy2020governance}.

  \begin{figure*}[t]
  \centering
  \includegraphics[width=\linewidth]{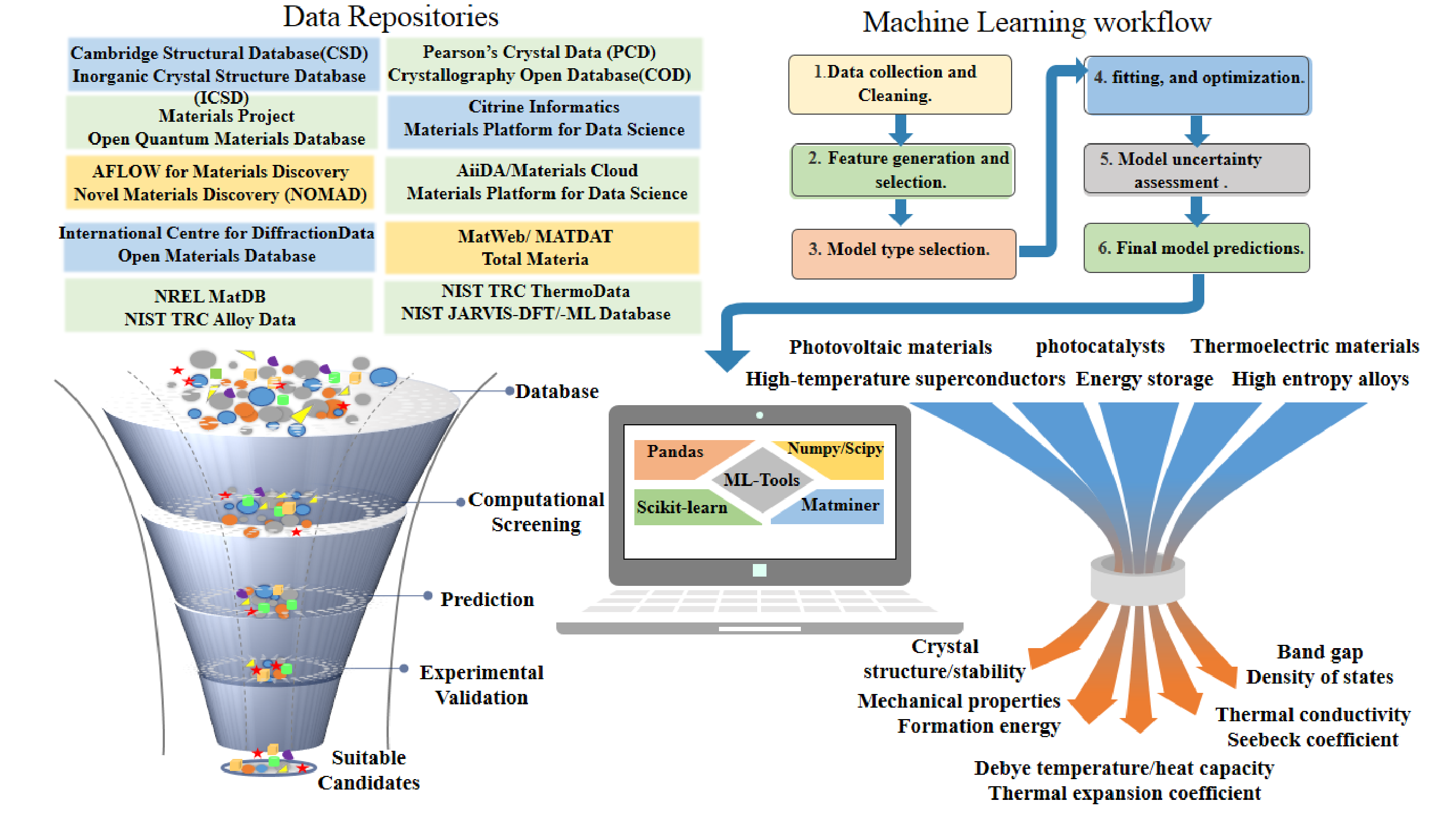}
   \caption{Schematic representation of a machine learning approach for finding high throughput materials. This diagram illustrates how various data repositories, machine learning tools, and processes work together to accelerate the discovery of novel materials.} 
     \label{fig:ML-GA}
\end{figure*}
At its core, ML involves the use of algorithms that can learn from data and improve over time, without requiring explicit programming. The importance of ML lies in its ability to process and analyze vast amounts of data and identifying patterns as well as anomalies that may not be apparent to humans. This makes it a valuable tool in many industries, from healthcare to finance to manufacturing, where it can be used to optimize processes, improve decision$-$making, uncover new business opportunities, and identifying materials with superior properties.
However, while ML has the potential to revolutionize many aspects of modern life, it is not a simple process. Developing effective ML models requires careful consideration of data quality, relevance, and cleanliness. In order to leverage ML, one must first identify the right data sources that can be applied to the learning process. This may involve collecting data from disparate sources and integrating them into a unified data model. Once data has been collected, it must be prepared for analysis by cleansing and tidying it. This process involves transforming the data into a format that can be understood by ML algorithms, which often require numerical data rather than textual information. 

Materials science is an interdisciplinary field that focuses on the study of the properties, structure, and synthesis of materials. However, traditional methods of materials discovery and design are often time$-$consuming and costly, which has hindered accelerated progress in this field.
The use of ML has rapidly transformed materials science by providing a powerful tool for accelerating the design and discovery of advanced functional materials and optimizing their synthesis process.~\cite{bishara2023state,panchal2019machine,ramprasad2017machine,lee2016prediction,oliynyk2016high,butler2018machine}. 

 In materials science, ML algorithms can analyze vast amounts of data from experiments and/or simulations stored in databases to identify patterns and relationships those may be difficult to detect using traditional approaches.
The collaboration of ML and computational materials science offers numerous advantages. One of the most significant benefits is the ability to predict the properties of materials from a wide chemical space before they are synthesized. By predicting the properties of materials based on their chemical composition, crystal structure, and other properties, ML algorithms can significantly reduce the time and cost associated with experimental synthesis and testing of materials~\cite{morgan2020opportunities,suh2020evolving,liu2017materials}.
The ML algorithms can also aid in the discovery of new materials with desired properties. By analyzing extensive databases of materials, ML algorithms can identify materials with properties similar to the desired ones and propose modifications to optimize their properties. This approach has the potential to accelerate the materials discovery and design process~\cite{juan2021accelerating,cole2020design,oganov2019structure,pulido2017functional}.

There are several comprehensive materials databases available to researchers those provide access to an extensive list of data for materials, which is essential for the discovery, design, and optimization of new materials. Most of these databases support Application Programming Interface (API) based data retrieval, allowing users to efficiently mine, sort, and, retrieve specific data they need.
Web based applications such as aflowlib~\cite{curtarolo2012aflowlib}, JARVIS~\cite{choudhary2020joint}, novel materials discovery laboratory~\cite{draxl2019nomad} (NOMAD) encyclopedia, crystals AI~\cite{zheng2020random}, citrine informatics (thermoelectric predictor)~\cite{gaultois2016perspective}, NIMS tools~\cite{tanifuji2019materials}, SUNCAT~\cite{winther2019catalysis} catalysis property predictor, and matlearn~\cite{peterson2021materials} use advanced ML algorithms and artificial intelligence techniques to predict various properties of materials. These tools can help researchers 
to screen and design new materials with desirable properties, such as thermoelectric transport properties, electronic structure, catalytic activity, etc. 
JARVIS, crystals AI, thermoelectric predictor, NIMS tools, SUNCAT, and aflow library are materials simulation software and they use density functional theory (DFT) and ML algorithms to predict properties of materials. Inorganic crystal structure database (ICSD)~\cite{belsky2002new}, open crystal database (OCD), Pearson crystal structure data (PCXD), and Cambridge Structure Database (CSD) are some of the crystal structure databases those possess experimentally measured crystal structures data of materials. While the open quantum materials database (OQMD)~\cite{saal2013materials}  is a database of quantum mechanical calculations for materials properties. Materials project~\cite{jain2013commentary} is a web$-$based platform that provides materials data and computational tools for materials scientists and engineers. Citrine informatics is a materials informatics company that uses ML algorithms to analyze and predict material's properties. Some of the other databases are the materials web online data base~\cite{friedman2009learner}, computational materials repository~\cite{rasmussen2015computational}, materials cloud platform for two$-$dimensional (2D) materials~\cite{mounet2018two}, Harvard clean energy project (CEP) for organic photo$-$voltaics materials~\cite{hachmann2011harvard}, and thermoelectrics design lab for thermoelectric materials~\cite{gorai2016te}. All these tools and resources are important for materials scientists and engineers to design and discover new materials with desired properties. Each of these web applications specializes in predicting a specific set of material properties and they often include databases of material's properties those can be used to search for and compare materials with specific properties.

The use of ML has significantly decreased the computational time required for calculating materials properties, as compared to first principles DFT calculations~\cite{hohenberg1964inhomogeneous,kohn1965self}. Moreover, this has been achieved without compromising accuracy and has enabled fast descriptive and predictive learning. Further, ML is particularly useful in handling multidimensional data, which is critical for accelerating materials development. Photovoltaics, piezoelectrics, magnetocalorics, magnetoelectrics, and thermoelectrics are some functional materials systems those have benefited from this revolution with demonstrated success in screening and designing efficient materials using ML algorithms. In materials science, selecting the most suitable ML algorithm for a particular application in a specific domain can also be a challenging task. This is because, different algorithms are designed to solve different types of problems and the effectiveness of a particular algorithm can depend on the specific characteristics of the materials data and the features used to train the model. One important consideration when selecting a ML algorithm is the type of materials data that is being analyzed. For example, if the data involves identifying patterns in X$-$ray diffraction spectra or crystallographic data, then ML algorithms such as clustering and pattern recognition may be appropriate. On the other hand, if the data involves predicting mechanical properties or optimizing material processing conditions, then regression and optimization algorithms may be more suitable. If the database has a huge number of data for a specific property then more advanced deep learning techniques may be suitable to employ.

Another factor to consider when selecting a ML algorithm in materials science is the size and complexity of the data. For instance, in the case of high$-$throughput materials screening, where large volumes of data are generated from diverse materials properties, unsupervised learning algorithms such as principal component analysis (PCA) or clustering may be useful for reducing the dimensionality of the data and identifying important features. Additionally, the desired outcome of the analysis is also a crucial factor in selecting an appropriate ML algorithm. For example, if the goal is to identify new materials with specific properties, then unsupervised learning algorithms such as clustering or generative adversarial networks (GANs) may be more useful~\cite{tshitoyan2019unsupervised,jang2020structure,graser2018machine,wei2019machine}. Finally, the computational resources available and the level of interpretability required are also important considerations when selecting a particular ML algorithm in materials science. Deep learning algorithms such as convolutional neural networks (CNNs) and recurrent neural networks (RNNs) may be more computationally expensive, but they can often provide highly accurate predictions. On the other hand, simpler algorithms such as Decision tree (DT) or Linear regression may be more interpretable and easier to implement. Anyway, whatever model we adopt, the quality of the dataset and appropriate features are key to predict the properties reliably. Here we have adopted regression methods because they are relatively less complex and easy to interpret and evaluate. The motivation behind selectively choosing the models used in this study is their proven superiority in performance on various datasets from diverse fields.
\par
 Priyanga \textit{et al.}~\cite{mattur2022prediction} compared the performance of different ML classifier models for predicting the nature of bandgap. She compared Logistic regression, K$-$Nearest Neighbors (KNN), Support vector clustering,
Random Forest (RF)~\cite{breiman2001random,liaw2002classification,svetnik2003random,li2013using}, DT~\cite{kushwah2022comparative}, Light Gradient Boosting Machine (LGBM)~\cite{ke2017lightgbm} and eXtreme Gradient Boosting (XGB)~\cite{chen2016xgboost,chen2015xgboost} for their predictability of  nature of bandgap for perovskite materials.
Wang \textit{et al.} trained a stacking model~\cite{pavlyshenko2018using,meharie2022application} for bandgap value prediction with LGBM, RF, XGB, and Gradient Boost Decision Tree (GBDT) as base models~\cite{wang2022accurate}.
Further ML models such as CatBoost~\cite{prokhorenkova2018catboost}, Gradient Boosting (GB)~\cite{friedman2001greedy,friedman2002stochastic,guelman2012gradient}, HistGradientBoosting, ExtraTrees, XGB, DT,
bagging~\cite{breiman1996bagging,buhlmann2012bagging,sutton2005classification,friedman2007bagging}, LGBM, GaussianProcess, artificial neural network (ANN), and light long short$-$term memory (LightLSTM) were compared for predictability of photocatalytic degradation of malachite green dye~\cite{jaffari2023machine}.
These are few latest studies on material properties using ML models.
\par
The bandgap is a fundamental material property that plays a critical role in different optoelectronic applications. The range of bandgap value required depends on the application and also different materials with varying bandgap values are necessary for various purposes. Materials with a small bandgap value are useful in electronics and detectors, while materials with a large bandgap value are useful in solar cells, LED, and transparent conducting applications. Materials with a narrow bandgap value are useful in thermoelectric devices. Estimating the bandgap value accurately is crucial in Materials Science and can lead to the development of new materials with enhanced properties for various applications, especially, for green energy technologies. Predicting the bandgap value of materials is an important task in Materials Science and has been the subject of many published papers~\cite{elsheikh2014review,radecka2008importance,okumura2015roadmap,takahashi2007wide}.
To choose the correct model for bandgap prediction, it is important to consider several factors, including the choice of ML algorithm, the quality and size of the dataset, feature selection, model validation, and model interpretability.	
There are many ML algorithms those can be used for predicting bandgap values, including regression models and neural networks. Each algorithm has its own strengths and weaknesses and the choice of algorithm will depend on the characteristics of the dataset as well as the specific problem being addressed. The quality of the data used to train the model is critical to its accuracy and generalizability.
\par
The another important aspect is that the dataset should be large enough to capture the full range of variability in the material properties and should be representative of the materials being studied to avoid biases. The dataset should also be clean and free of errors or outliers those could skew the predicting capability. The selection of features or inputs those are used to predict bandgap value are important factors in determining the accuracy of the model. Relevant features may include elemental composition, crystal structure, and other material properties. It is important to select features those are relevant to the problem being addressed and that have a strong correlation with the targeted property.  Once a model has been trained with the datasets and appropriate features, it is important to validate its performance on a separate test dataset that has not been used for training. This helps to ensure that the model is able to generalize to new data and is not overfitting to the training dataset. The ability to interpret the results of a model is also important for understanding the factors those are driving the predictions. It may be noted that, the models those are more interpretable, such as DT or Linear regression models, can provide insights into the underlying relationships between the input features and the targeted property.
\par
In their study, Heng \textit{et al.} utilized neural networks and leveraged available experimental data from the literature to predict the bandgap of binary semiconductor alloys~\cite{heng2000prediction}. They focused on the elemental properties of the constituents and observed a continuous variation of bandgap values in A$_i$B$_{1-i}$ alloys as the mole fractions of the binary constituents changed. This observation led them to conclude that the bandgap value is influenced by the interplay between attractive and repulsive forces of the constituents as well as the diffusion rate of atoms during the formation of the semiconductor.

Zhuo \textit{et al.} employed variety of ML methods to model the bandgap energies of 3896 inorganic solids those were reported in experimental studies~\cite{zhuo2018predicting}. Their analysis covered a wide range of materials and aimed to accurately predict their bandgap values. By incorporating different ML techniques they sought to improve the understanding and prediction of bandgap values in diverse inorganic solids. Zeng \textit{et al.}~\cite{zeng2002prediction} extended this analysis to ABC$_2$ type chalcopyrites. They employed neural network techniques to explore the relationship between elemental properties and bandgaps values in this class of materials. By investigating a different class of compounds, they aimed to expand the understanding of the factors influencing bandgap energies.
In a different approach, Weston and Stampfl~\cite{weston2018machine} combined DFT calculations with multiple ML models to predict the bandgap value of Kesterite I$_{2}$-II-IV-V$_{4}$ semiconductors. They trained their models using 200 bandgap values calculated using the hybrid HSE functional and focused on classifying these bandgap values as either direct or indirect. This approach allowed them to gain insights into the bandgap characteristics of the Kesterite materials.

These studies collectively demonstrate the application of neural networks, DFT calculations, and ML models in predicting and understanding the bandgap behavior of various semiconductor alloys and inorganic solids. Each study contributed to advancing the knowledge in this field and provided valuable insights into the factors influencing bandgap value and its nature. However, while ML holds enormous potential, it is not a simple process. One of the biggest challenges is collecting and preparing data for analysis. In order to train a ML model, one should first identify the right data sources and ensure that the data is accurate, relevant, and clean. This may involve combining data from multiple sources, such as open databases, literature's etc and transforming them into a format that can be understood by ML algorithms. Another challenge is selecting the right ML algorithm for a given task. There are many different algorithms to choose from, each with its own strengths and weaknesses. Some algorithms are better suited for classification tasks, while others are better suited for regression tasks. Some algorithms are more robust to noise and outliers, while others are more sensitive to them. Selecting the right algorithm requires careful consideration of the data and the problem at hand. Despite these challenges, ML has become an essential tool for one looking to leverage their data assets and gain a deeper understanding of their task and to accelerate it. By following best practices for data preparation and algorithm selection, one can develop robust and accurate ML models that drive innovation and create new opportunities.

\section{Data Preparation and Feature Engineering }\label{Data preparation, feature engineering}

In this section, we will provide a comprehensive discussion on how to achieve accurate predictions for the bandgap values for HH compounds. We would like to stress the importance of feature engineering in ML methods, as it can significantly affect the predicting capability of the ML models. The selection and engineering of relevant features for a given dataset are crucial for achieving the best possible model accuracy. However, featurizing and selecting specific features with good correlation with targeted properties for a dataset can be a challenging and time$-$consuming task. Therefore, we aim to use easily accessible information with the input parameters as simple as possible. We have used the featurization method provided in matminer based on both the composition and structure of the materials, We have used 8 ML models based on the selection of input features, which we will explain in detail in the following sections.

\begin{figure*}[t]
  \centering
  \begin{subfigure}[a]{0.7\textwidth}
    \includegraphics[width=\linewidth]{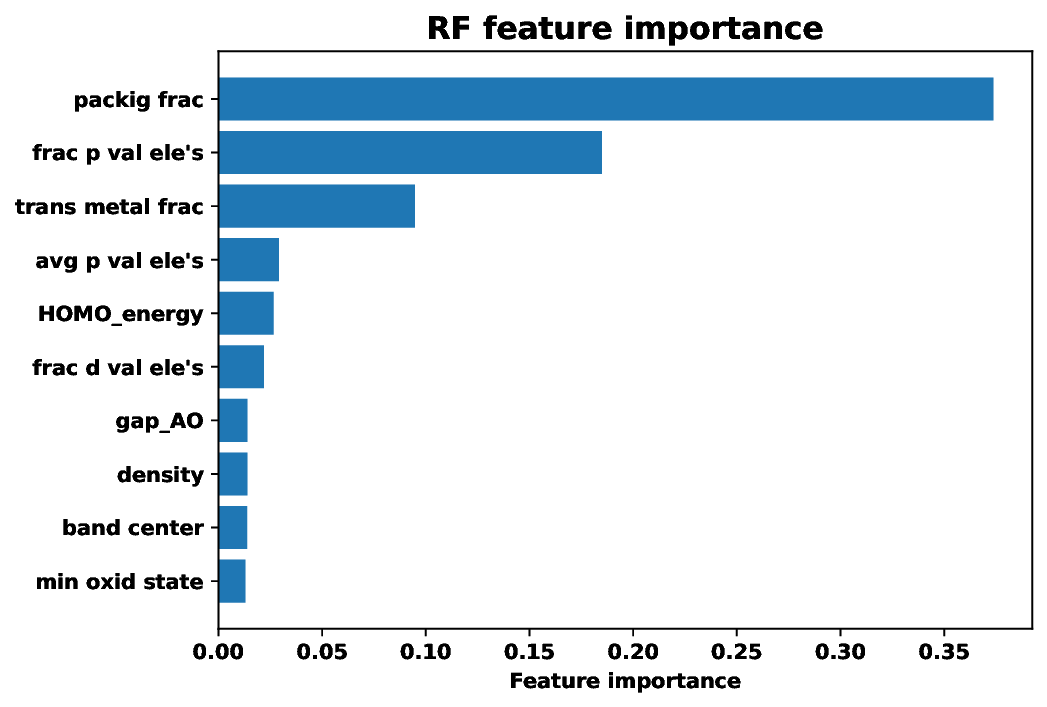}
  \end{subfigure}
  \hfill
  \begin{subfigure}[b]{0.5\textwidth}
    \includegraphics[width=\linewidth]{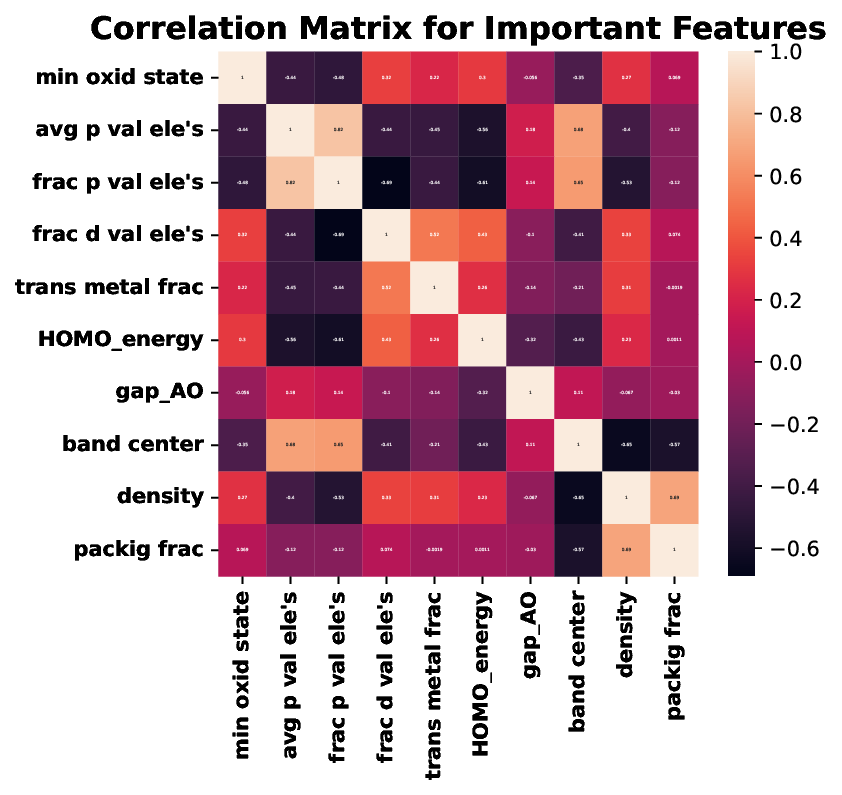}
    \end{subfigure}
   \caption{(a)  Primary feature importance with the bandgap values obtained from Random forest regression (RF) model (b) statistical heat map showing correlation between primary features used to train the RF model
} 
\label{fig:RF}
\end{figure*}

Our objective is to develop appropriate ML models those  can accurately predict the bandgap values of large number of HH compound compositions derived based on empirical 18 VEC rule those are expected to be semiconductors, which is essential for identifying new materials with desired bandgap value. We can determine the most effective set of input features for predicting the bandgap values of HH compounds. Through our analysis and discussion, we hope to provide valuable insights into the process of accurately predicting the bandgap values of HH compounds using ML techniques and it can be generalized to predict bandgap values for all inorganic solids. To identify the most relevant features, we used the permutation of features for feature importance and correlation matrix heatmap analysis, which is shown in Fig.\ref{fig:RF}. To assess the performance accuracy of each model, two standard performance metrics were considered and those are RMSE (Root Mean Square Error) and  R$^2$ (coefficient of determination). Among these two performance matrices, the R$^2$ metric is a dimensionless statistical measure that indicates how well the data fits a regression line, typically ranging from 0 to 1 and RMSE provides a measure of the average deviation from the actual value. The RMSE expresses the average model predictions in the units (here eV) of the targeted property (here bandgap) used to train the data and is a negatively oriented score, meaning that a lower score indicates better performance of the model. These two performance matrices are defined as.

\begin{equation}
RMSE=\sqrt{\frac{1}{N}\sum_{i=1}^{N} \left( y_i - \hat{y}_i\right)^2}
\end{equation}
and
\begin{equation}
R^2=1-\frac{\sum_{i=1}^{N} \left( y_i - \hat{y}_i\right)^2}{\sum_{i=1}^{N} \left( y_i - \bar{y}_i\right)^2}
\end{equation},
where  y$_i$ represents the true value, $\hat{y}_i$ represents the predicted value, and $\bar{y_i}$ represents the average value. Additionally,$i$ ranges from 1 to N, where N represents the total number of targeted property data present in the test dataset.

To enhance the quality of our feature set, we employed the Pearson correlation method using the pandas library. This technique allowed us to identify features closely related to our targeted property. We then examined a heatmap illustrating the Pearson correlation coefficients (PCC). Only PCC values surpassing the absolute threshold of 0.2 were visualized, as shown in Figure \ref{fig:pcc}. These values ranged from -1 to 1, with a higher absolute value indicating a stronger correlation between the target and features. Analyzing the PCC heatmap, we were able to eliminate some features with little or no correlation to the targeted property. For example, a feature with a PCC value close to 0 was likely to be removed, as it would contribute minimally to the model's predictive ability. Conversely, features with high PCC values were retained, as they were likely to be important predictors of the output variable. The correlation matrix among features as shown in figure\ref{fig:RF}b is also obtained using the Pearson correlation method, which calculates pairwise correlations between features in the columns of a data frame. The resulting correlation values range from -1 to 1, with a higher absolute value signifying a stronger correlation between features. By analyzing the correlation heat map, we effectively eliminated highly correlated features mitigating issues like multicollinearity and overfitting. Through these steps, we reduced the dimensions of our feature space while maintaining its informative aspects.

Our feature selection and processing methods played a crucial role in developing our predictive models. They helped us identify critical features and thus enhance the model's predictive capacity.

\begin{figure*}[t]
  \centering
    \includegraphics[width=\linewidth]{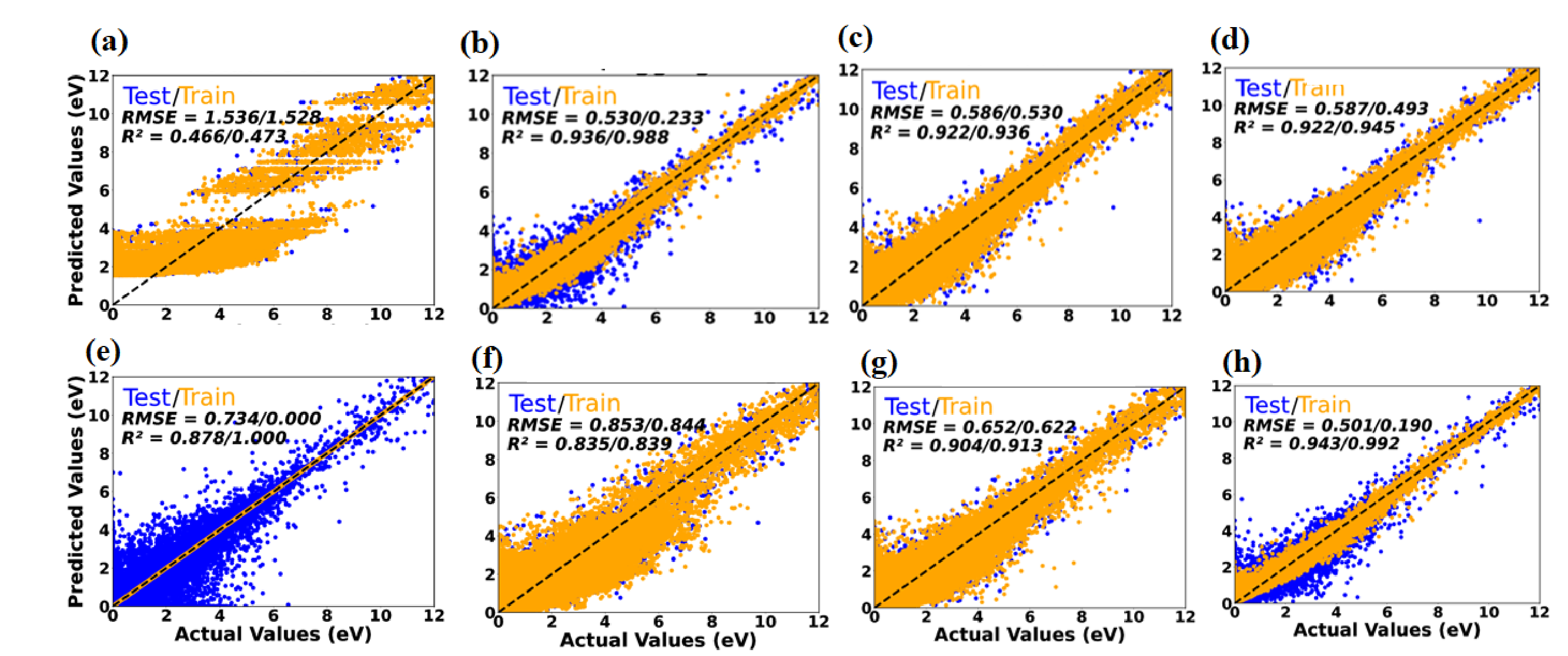}
     \caption{Learning performance of (a) AdaBoost (b) bagging  (c) CatBoost (d) XGB   (e) decision tree (f) Gradient Boosting (g) LightGBM (h) Random Forest.} 
     \label{fig:8model}
\end{figure*}
\section{Machine Learning Algorithm} \label{Machine Learning Algorithm}

Composition dependent descriptors are a type of material properties those describe how the properties of a material depend on its composition or the relative amounts of the different elements in the material. One such descriptor is the stoichiometric average of the elemental properties, which is a weighted average of the elemental properties based on their stoichiometry in the material. The stoichiometric average is a useful descriptor because,  it takes into account the contributions of all the elements in the material, rather than just one or two dominant elements. The elemental properties those are typically used to calculate the stoichiometric average include the atomic number, ionic radius, electronegativity, and electron configuration of the constituent elements. To calculate the stoichiometric average, the elemental properties are first weighted by their stoichiometric coefficients, which are the numbers those represent the relative amounts of each element in the material. The stoichiometric average can be used to predict the properties of materials based on their composition. For example, it has been found that the stoichiometric average of the electronegativity of the elements in a material can predict its bandgap value if it originates from the ionicity, which is an important property for electronic and optoelectronic applications. Other properties those can be predicted using the stoichiometric average include the thermal and mechanical properties of materials.

We have used the matminer library to obtain a dataset of materials with known bandgap values obtained from GGA$-$PBE calculations and their corresponding composition as well as structure features. In addition to these features, we incorporated the mean, maximum, and standard deviation of atomic numbers as well as the structural features of the corresponding materials by accounting the GGA$-$PBE calculated lattice parameter values $a$, $b$, $c$, $\alpha$, $\beta$ and $\Gamma$ obtained from the corresponding calculated crystal structure.

We have split the generated dataset into a training set and a testing set with 8:2 ratio. The training set was used to train our models and the testing set was used to evaluate their performance. We have employed the following eight regression models such as  RF, GB, XGB, LGBM, DT, AdaBoost~\cite{freund1996experiments,solomatine2004adaboost,collins2002logistic,shrestha2006experiments} to identify the best performing model. For the performance analysis, we have considered R$^2$ and RMSE performance metrices to select the best algorithm for our dataset. After assessing the performance of the considered regression models on test data, we found that the RF regressor exhibited the best performance among the considered ML models followed by that the bagging regressor performed well. We have also considered the following seven models as a base for the bagging ensemble model analysis such as  DT, CatBoost, GB, LassoCV~\cite{tibshirani1996regression}, LGBM, RidgeCV, XGB With AdaBoost as booster. Among the various models considered in the bagging ensemble model analysis, we found that the bagging with DT and AdaBoost showed the best performance. 

The ML regression models were trained using available bandgap values from matminer database obtained from DFT calculations based on GGA$-$PBE functional , excluding the well studied compounds those such as Si, GaAs, TiO$_2$ etc were used for validation. Subsequently, we have performed band structure calculations using the GGA$-$PBE functional to cross validate the ML predicted bandgap value of newly identified HH compounds. Further improve the DFT predicted  bandgap value we have also used the TB$-$mBJ functional and the corresponding calculated bandgap values are compared in table\ref{tab:stacking_models}. This combination of ML and DFT  hybrid approach holds promising potential for accelerated and reliable bandgap predictions in new semiconducting materials.

\begin{table*}
\centering
\caption{Model performance on different datasets (test/train)}
\label{tab:models}
\small
\setlength{\tabcolsep}{3pt}
\begin{tabular}{|l|c|c|c|c|}
\toprule
\multicolumn{5}{|c|}{\textbf{Datasets}} \\
\hline
Model & Dataset(1) 9223 & Dataset(2) 23173 & Dataset(3) 46528 & Dataset(4) 94815 \\
\hline
Adaboost & $0.521/0.524$ & $0.327/0.361$ & $0.361/0.375$ & $0.466/0.473$ \\
& 1.252/1.248 & 1.429/1.420 & 1.502/1.493 & 1.536/1.528 \\
\hline
Bagging & $0.859/0.971$ & $0.856/0.976$ & $0.901/0.981$ & $0.936/0.988$ \\
& 0.680/0.306 & 0.662/0.277 & 0.591/0.260 & 0.534/0.235 \\
\hline
CatBoost & $0.877/0.955$ & $0.868/0.93$ & $0.896/0.925$ & $0.922/0.936$ \\
& 0.636/0.385 & 0.634/0.470 & 0.607/0.519 & 0.586/0.530 \\
\hline
LGBM & $0.878/0.946$ & $0.856/0.909$ & $0.878/0.898$ & $0.904/0.913$ \\
& 0.631/0.419 & 0.660/0.535 & 0.657/0.602 & 0.652/0.622 \\
\hline
Random Forest & $0.873/0.984$ & $0.872/0.984$ & $0.91/0.987$ & $0.943/0.992$ \\
& 0.620/0.225 & 0.623/0.226 & 0.563/0.213 & 0.501/0.190 \\
\hline
Decision Tree & $0.729/1$ & $0.756/1$ & $0.809/1$ & $0.878/1$ \\
& 0.942/0.0 & 0.860/0.0 & 0.822/0.0 & 0.734/0.0 \\
\hline
GB & $0.826/0.851$ & $0.771/0.809$ & $0.789/0.808$ & $0.835/0.839$ \\
& 0.755/0.697 & 0.833/0.776 & 0.844/0.827 & 0.853/0.844 \\
\hline
XGB & $0.877/0.981$ & $0.867/0.955$ & $0.891/0.940$ & $0.922/0.945$ \\
& 0.635/0.248 & 0.636/0.376 & 0.619/0.462 & 0.587/0.493 \\
\hline

\end{tabular}
\end{table*}


\section{Results and Discussion}
\subsection{Training  ML Algorithms to Create an Accurate ML model}

The calculation of properties of materials is a complicated and computationally intensive process. 
The DFT based computational methods are commonly used to calculate almost all the properties of materials. However, when attempting to discover materials with certain desirable properties, this approach can be time$-$consuming and computationally expensive. To address this issue, ML approaches based on appropriate features that describe material properties can be developed. Typically in feature engineering, features those strongly correlated with targeted properties are selected while weakly correlated features are ignored in the materials feature set by the feature elimination process. This is because, incorporating weakly correlated features can increase computational complexity and potentially decrease the accuracy of the model. Currently, ML$-$based materials property prediction mainly relies on composition and geometric structural properties based features. Also, the accuracy of the predicted results heavily rely on the quantity and the quality of the input dataset.
 
Figure \ref{fig:RF} displays a heatmap presenting the PCC values of the most significant features used for training our RF ML regressor such as packing fraction, fraction of $p$ valence electrons, transition metal fraction, average of $p$ valence electrons, HOMO energy etc. This model is found to be the best performer among the eight stand-alone ML regression models examined in the current study. A comprehensive list of selected features used for these models is provided as supplementary information.

The heatmap shown in Figure \ref{fig:RF} was generated for the RF model using the Pearson correlation coefficient (PCC) values obtained from dataset 4. The colors on the heatmap illustrate the correlations between different features. Strong positive correlations are shown with less intense red colors, while strong negative correlations are represented by intense red colors. The diagonal elements are always 1, indicating how each feature correlates with itself. By looking at the correlation heatmap one can identify the important features those are highly correlated. For example, the average number of $p$ valence electrons and the fraction of $p$ valence electrons in constituent elements in the compound are highly positively correlated. However, the band center and packing fraction have a negative correlation as evident from Fig.\ref{fig:RF}.

\begin{figure*}
    \centering
    \begin{subfigure}[b]{0.48\textwidth} 
        \centering
        \caption{ }
        \includegraphics[height=6cm]{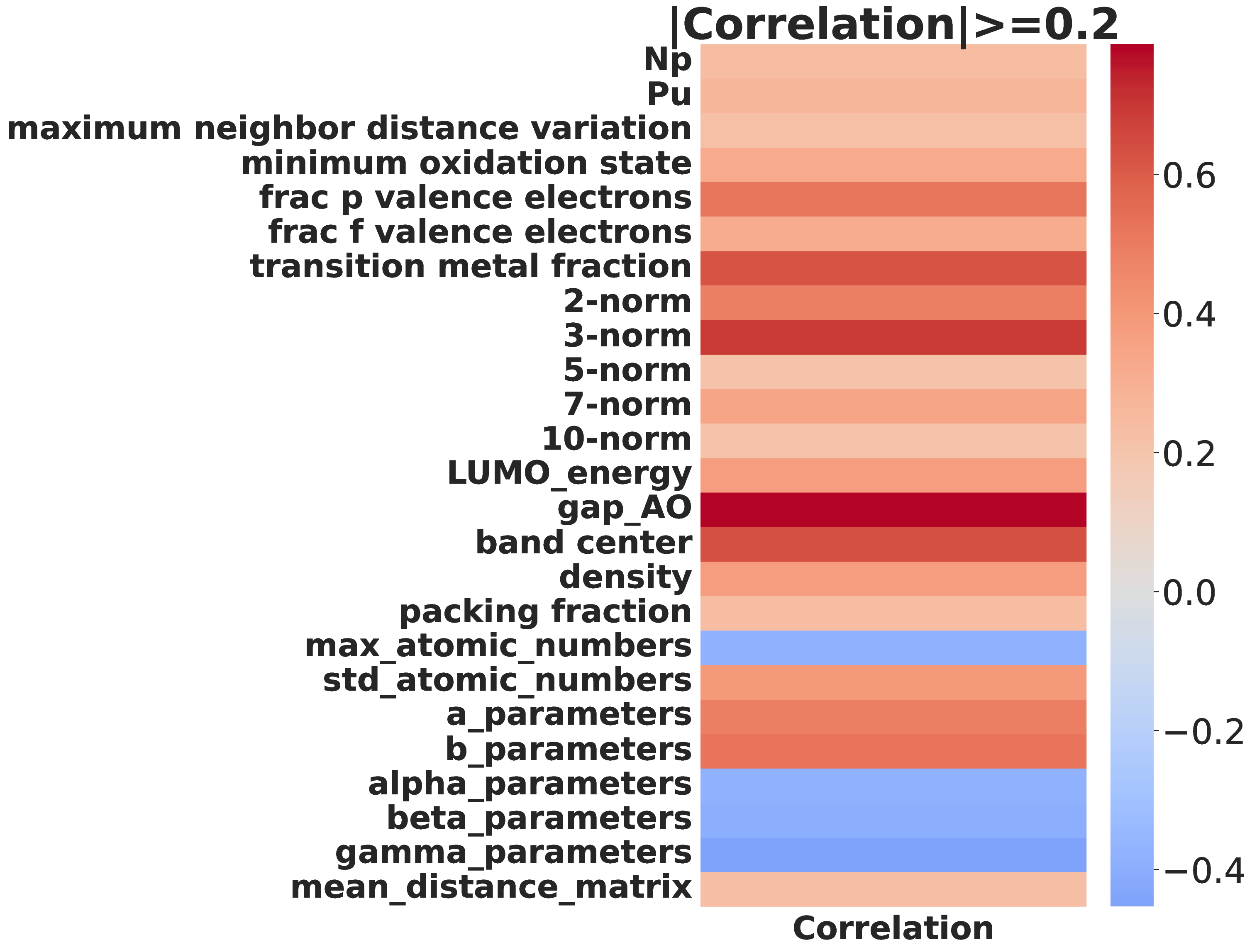}
        \label{fig:pcc}
    \end{subfigure}\hspace{0.001\textwidth} 
    \begin{subfigure}[b]{0.48\textwidth} 
        \centering
        \caption{ }
        \includegraphics[height=5.8cm]{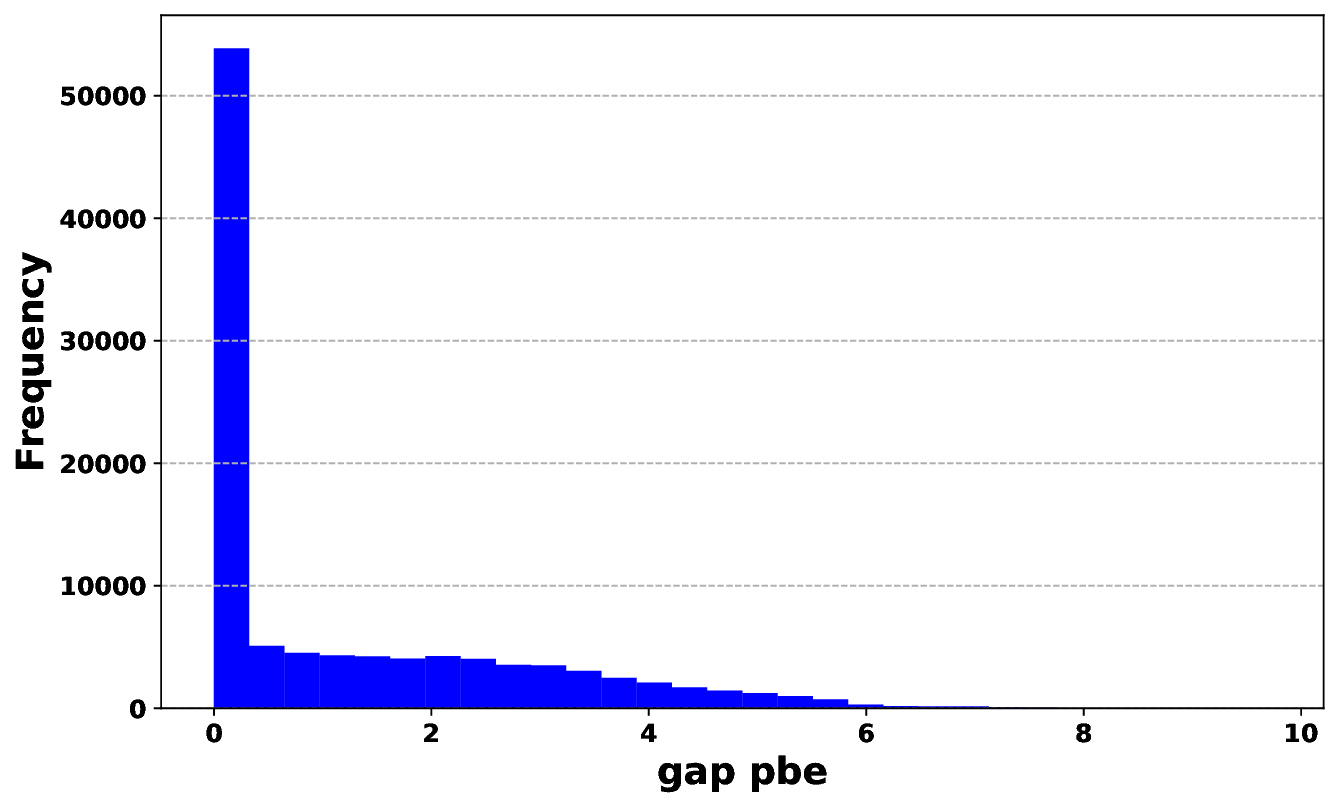} 
        \label{fig:gau}
    \end{subfigure}
    \caption{(a) Targeted property vs feature correlation heatmap and (b) distribution of bandgap values}
\end{figure*}

Apart from analysing the correlation between features we also analysed the correlation between features and the targeted property. In the heatmap for target property-features correlation shown in Fig.\ref{fig:pcc}, weak negative correlations are shown in blue colour while shades of red colour represent the strength of positive correlations. Both heatmaps may help in the development of more accurate prediction models by providing insights into the physical mechanisms through various features those determine the bandgap values of materials. A comparison of the performance of 8 ML regression models including AdaBoost, bagging, CatBoost, LGBM, RF, DT, GB, and XGB, with various dataset sizes mentioned in section \ref{Machine Learning Algorithm}  are presented in Table\ref{tab:models}. By comparing different models across various dataset sizes, we have made several observations as follows.

For the performance analysis, we started by examining the AdaBoost regression model. The results indicate that, when tested on 20\% of dataset 1, the AdaBoost regression model yielded an $R^2$ value of 0.521 and a  RMSE value of 1.252. For the ML model used for the above case we have trained with  80\% of dataset 1 for which we obtained $R^2$ and RMSE values of 0.524 and 1.248 eV, respectively. However, as we increased the dataset size, we observed a slight decline in the model's performance. This is evident from the decreasing $R^2$ values and increasing RMSE values when we trained the ML model with the dataset 2, 3 and 4. This reduced performance is likely attributed to the presence of imbalanced data, which is illustrated in Fig. \ref{fig:gau}. The majority of bandgap values are concentrated between zero and two. It may be noted that the AdaBoost regressor is less adoptable for handling imbalanced data compared to RF and XGB, as discussed earlier \cite{shahri2021comparing, krawczyk2016learning}.

Similarly, we conducted an assessment of the performance of other ML regression models with dataset 1, 2, 3 or 4. This evaluation encompassed an analysis of both RMSE and $R^2$ values for testing as well as training datasets. The outcomes of this evaluation are meticulously presented in Table\ref{tab:models}. It is worth noting that, the emphasis of this analysis lies in the model's performance on the test dataset, as it provides insight into the model's predictive capability on new/unseen data as we done with 20\% of unseen data present in the original dataset.

The ML regression models such as Bagging, CatBoost, LGBM, and RF demonstrated competitive performance on both smaller as well as larger dataset sizes. These models consistently displayed high $R^2$ values across various dataset sizes mention in section \ref{Machine Learning Algorithm}. The corresponding decrease in RMSE values indicate that their predictions were highly reliable. This suggests that these regressors can effectively capture the complex relationships between various features with targeted property in smaller datasets and further enhance their performance with the addition of more data. The RF regressor consistently demonstrated high $R^2$ values irrespective of the dataset used and relatively low RMSE values across all dataset sizes considered to train the model in the present study. In the case of dataset 1, the RF model achieved an $R^2$ value of 0.873 and an RMSE of 0.620 eV on testing, indicating its ability to capture a substantial portion of the variance in the targeted property, Moreover, its performance improved further as the dataset size increased. For dataset 4, RF model exhibited the best performance among all the stand-alone models we have analysed here with the R$^2$ and RMSE values  0.943 and 0.501 eV, respectively during testing.

In contrast, the DT regression exhibited limited predictive power when compared to other models. For example, in case of dataset 1, this model achieved an $R^2$ value of 0.729 and an RMSE value of 1.0 eV during the testing, signifying a relatively weak capacity to capture the variance in the targeted property. However, on the training set, it achieved an $R^2$ of 0.942 and an RMSE of 0 eV, implying that the model significantly  overfitting to  the training data. This inconsistency of DT model performance on the training and testing dataset suggests that it might have overfitted the training data, causing difficulties in generalizing its performance to the test dataset. An important conclusion can be drawn from the above observations that models those are particularly adopt to handle  data imbalances and capturing nonlinear relationships between features and also with features and targeted property tend to perform better. Notably, among these models, the RF stands out as particularly well suited for predicting the bandgap values more reliably. The scatter plots for individual models with averaged regression metrics and the predictions corresponding to the training as well as the testing datasets as inputs are shown by orange and blue circles, respectively in Fig. \ref{fig:8model}.

It is important to additionally consider how a specific model performs on training dataset so that one can evaluate  how well it predicts on test data. This enables us to determine the model's adaptability to various datasets and the presence or absence of overfitting or underfitting, which could have an effect on the algorithm's performance. When the data points from the training set align exactly with the dashed reference line, this is a  sign of overfitting. In this scenario, the algorithm has essentially "memorized" the training data, which might render it less capable of extrapolating to new and unseen data. Such rigidity could undermine the algorithm's accuracy/predicting capability for generalised dataset when tasked with predicting bandgap values for previously unencountered materials. Conversely, if the training data points significantly deviate from the dashed line, it points to the presence of underfitting. Underfitting suggests that the algorithm hasn't fully captured the underlying patterns and complexities as well as nonlinearity of the data, resulting in a less accurate predictive capacity. This situation can lead to sub-optimal performance when applied to new datasets, as the algorithm lacks the capacity to generalize effectively. This means that, a balanced algorithmic performance is needed in both training and testing datasets to reliably predict the properties and obviously for any unseen new data. This balance helps to guarantee that the algorithm can forecast GGA$-$PBE bandgap values for a wide range of materials with high accuracy. The study presented here opens the path for identifying appropriate algorithm for targeted  property prediction model development with improved the accuracy.

\begin{figure*}[t]
  \centering
    \includegraphics[width=\linewidth]{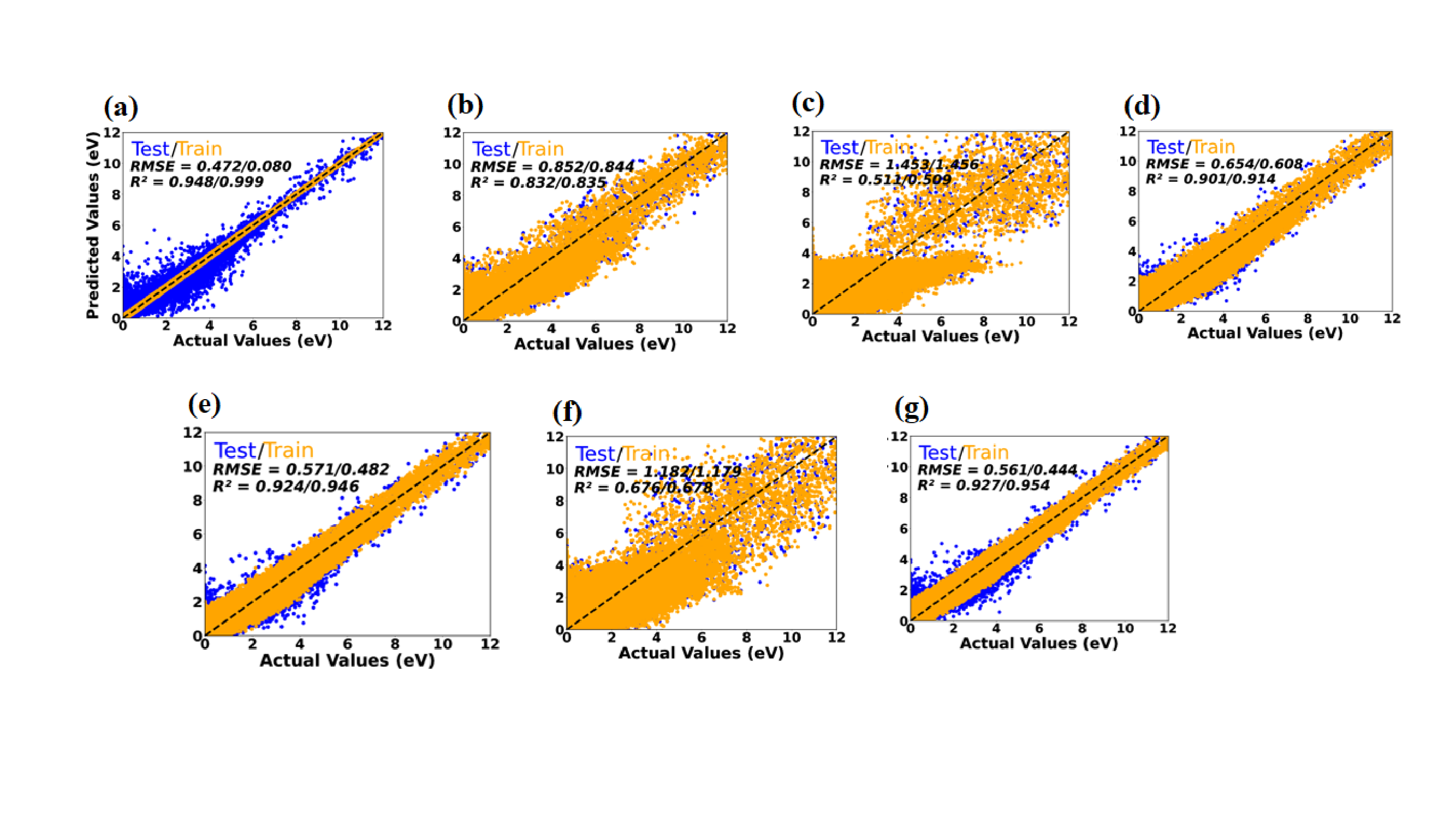}
    \vspace{-2cm}
   \caption{Scatter plots for various bagging model performance with AdaBoost as a booster and (a) DT, bagging, and AdaBoost (b) GB, bagging, and AdaBoost (c) LassoCV, bagging, and AdaBoost (d) LGBM, bagging, and AdaBoost  (e) Catboost, bagging, and AdaBoost  (f) RidgeCV, bagging, and AdaBoost and (g) XGB, bagging, and AdaBoost as base models.} 
\label{fig:bagging}
\end{figure*}

\begin{figure*}[t]
  \centering
    \includegraphics[width=\linewidth]{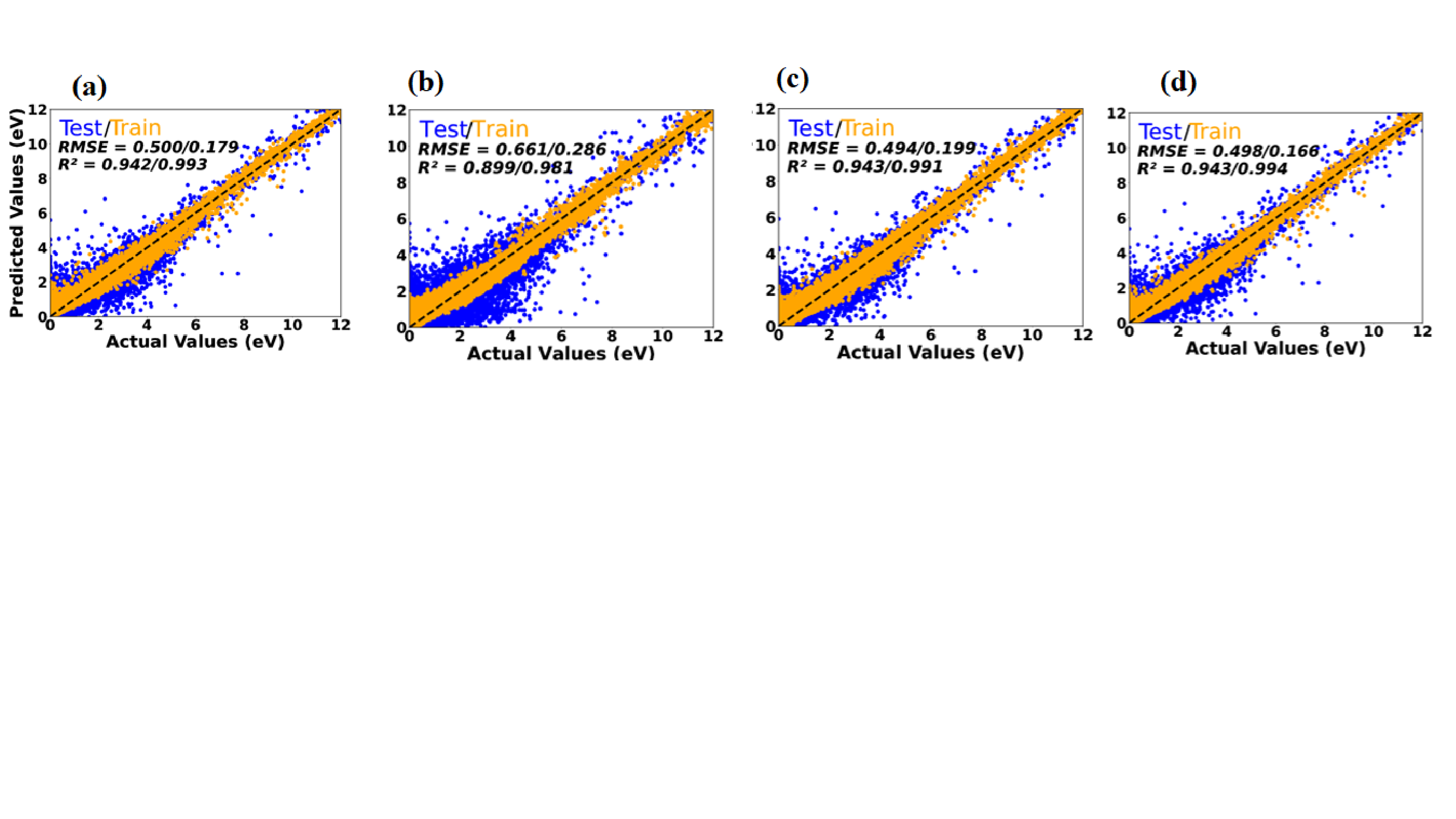}
    \vspace{-5.5cm}
   \caption{Scatter plot for stacking model performance  (a) RidgeCV, LGBM, RF as base model and LassoCV as meta$-$model (b) SVR, DT, KNN as base model and as GB meta$-$model (c) LGBM, RF, XGB as base model and RidgeCV as meta$-$model and  (d) Linear, RF, GB as base model and GB as meta$-$model.} 
\label{fig:stac}
\end{figure*}

\begin{table}[t]
\small\addtolength{\tabcolsep}{-2.7pt}
\centering
\caption{The performance of various bagging models }
\label{tab:bagging_models}
\begin{tabular}{lcc}
\hline
\textbf{Models} & \textbf{R\textsuperscript{2}} & \textbf{RMSE (eV)} \\
   & \textbf{Test/Train } & \textbf{Test/Train}\\
\hline
DT, bagging and AdaBoost   & 0.948/0.999   & 0.472/0.080   \\
Catboost, bagging and AdaBoost   & 0.924/0.946  & 0.571/0.482    \\
GB, bagging and AdaBoost  & 0.832/0.835   & 0.852/0.844   \\
LassoCV, bagging and AdaBoost & 0.511/0.509 & 1.453/1.456   \\
LGBM, bagging and AdaBoost   & 0.901/0.914  & 0.654/0.608    \\
RidgeCV, bagging and AdaBoost & 0.676/0.678 & 1.182/1.179  \\
XGB, bagging and AdaBoost  & 0.927/0.954  & 0.561/0.444 \\
\hline
\end{tabular}
\end{table}

\begin{table}[t]
\small\addtolength{\tabcolsep}{-3.5pt}
\centering
\caption{The performance of various stacking models}
\label{tab:stacking_models}
\begin{tabular}{lccc}
\hline
\textbf{Models} & \textbf{Meta$-$Models } & \textbf{R\textsuperscript{2}} & \textbf{RMSE (eV)} \\
& & \textbf{Test/Train} & \textbf{Test/Train}\\ 
\hline
LGBM, RF, XGB & RidgeCV & 0.943/0.991 & 0.494/0.199  \\
RidgeCV, LGBM, RF & LassoCV & 0.942/0.993 & 0.500/0.179  \\
SVR, DT, KNN, & GB & 0.899/0.981 & 0.661/0.286   \\
Linear, RF, GB, & GB & 0.943/0.994 & 0.498/0.166  \\
\hline
\end{tabular}
\end{table}

\begin{table*}[t]
\centering
\caption{Comparison of bandgap values(in eV) predicted by bagging and stacking models and the calculated bandgap value obtained from GGA$-$PBE as well as TB$-$mBJ based density functional calculations for newly predicted half Heusler alloys.
The bandgap values taken from earlier studies are given in brackets.}
\label{tab:stacking_models}
\begin{tabular}{|l|c|c|c|c|}
\hline
\textbf{Composition} & \textbf{Bagging model} & \textbf{Stacking model} & \textbf{GGA$-$PBE} & \textbf{TB$-$mBJ} \\
\hline
Si & 1.18 & 1.25 & 1.28~\cite{jain2013commentary}& \\
C & 2.75 & 3.015 & 4.11~\cite{jain2013commentary}& \\
SiC & 2.09 & 2.03 & 2.30~\cite{jain2013commentary}& \\
LiF & 7.68 & 8.08 & 8.70~\cite{jain2013commentary}& \\
BN & 4.70 & 4.532 & 5.77~\cite{jain2013commentary}& \\
TiO$_2$ & 2.89 & 3.20 & 3.42~\cite{jain2013commentary}& \\
GaS & 1.743 & 1.70 & 2~\cite{jain2013commentary}& \\
SiGe & 0.74 & 0.64 & 0.51~\cite{jain2013commentary}& \\
SiSn & 0.54 & 0.41 & 0.43~\cite{jain2013commentary}& \\
 LuSiAu & 0.10 & 0.07 & 0.33 & 0.43 \\
ScGeAu & 0.13 & 0.12 & 0.13 & 0.56 \\
YCoTe & 0.46 & 0.53 & 0.88 & 1.00 \\
ScSnAu & 0.13 & 0.10 & 0.12 & 0.40 \\
TiGaAg & 0 & 0.01 & 0.2 & 0.11 \\
LaNiBi & 0.20 & 0.20 & 0.34 & 0.55 \\
LuNiSb & 0.19 & 0.16 & 0.18 & 0.19 \\
ZrFeTe & 0.99 & 0.92 & 1.15 & 1.21 \\
ZrSbRh & 0.99 & 0.82 & 1.18 & 1.25 \\
LiZnSb & 0.19 & 0.11 & 0.53 & \\
LuNiAs & 0.15 & 0.12 & 0.45 & 1.37 \\
NbGaNi & 0.05 & 0.02 & 0.047 & 0.15 \\
LaNiSb & 0.12 & 0.23 & 0.45 & \\
LuSbPd & 0.14 & 0.07 & 0.33 & \\
CuSbS$_2$ & 0.81 & 0.78 & 0.72~\cite{jain2013commentary}& \\
LuSiAg & 0.16 & 0.13 & 0.62 & \\
LiZnAs & 0.32 & 0.43 & 0.35 & 1.37 \\
TiTeOs & 0.33 & 0.368 & 0.45 & 0.96 \\
TiFeTe & 0.67 & 0.77 & 0.98 & 1.2 \\
TiMnO$_3$ & 1.38 & 1.08 & 1.53~\cite{jain2013commentary}& \\
LiZnP & 1.15 & 1.17 & 1.34~\cite{jain2013commentary} & \\
LaNiP & 0 & 0.003 & 0~\cite{jain2013commentary}& \\
\hline
\end{tabular}
\end{table*}

To further improve the performance of our predicting capability of bandgap values through ML approach we have adopted emerging new ML methods such as bagging and stacking approaches. These ensemble ML regression models were utilized to reduce the variance in prediction and average out the predictions of models with similar bias values. A minimum benefit of using ensemble models is the reduction of average prediction spread. These methods are advantageous over 8 different stand-alone ML models used in the present study. Because, these ensemble models combine multiple individual models through a selection strategy, which results in greater generalisability and robustness than adopting a stand-alone ML model~\cite{odegua2019empirical}.

The bagging ensemble model, which stands for "Bootstrap aggregating", is a robust ensemble learning technique. It operates by dividing the training data into multiple subsets through a process called bootstrapping. Each of these subsets is used as input for individual base models. The results generated by these base models are then combined and fed into the AdaBoost model to produce the final prediction. In essence, bagging creates diversity among the base models and leverages their combined strength. In current study we have taken DT, CatBoost, GB, LassoCV, LGBM, RidgeCV and XGB as the base models and outputs from these base models are then feed into to the AdaBoost model for the final regression. 

From our detailed performance analysis we found that the DT taken as base model in bagging and AdaBoost as booster exhibited impressive performance, achieving high $R^2$ values of 0.948 and 0.999 for the test and train datasets, respectively. These results indicate that this model can effectively capture a significant portion of the variance in the targeted property. Furthermore, the corresponding test/train RMSE values of 0.472/0.080 eV suggest that this model's predictions are relatively close to the actual values with minimal error. The CatBoost as base model in this bagging model also demonstrated very good performance. It achieved R$^2$ values of 0.924/0.946 for the test/train datasets indicating its reliability to predict the bandgap values. This bagging model with Catboost as base accuracy was further validated by the relatively low RMSE values of 0.571/0.482 eV, indicating small prediction errors. The GB as a base model gave an average performance, achieving R$^2$ values of 0.832/0.835 and relatively high RMSE values of 0.852/0.844 eV for the test/train datasets, respectively.

\begin{figure*}
\centering
\begin{subfigure}[b]{\textwidth}
         \centering
         \caption{ }
         \includegraphics[height=8cm]{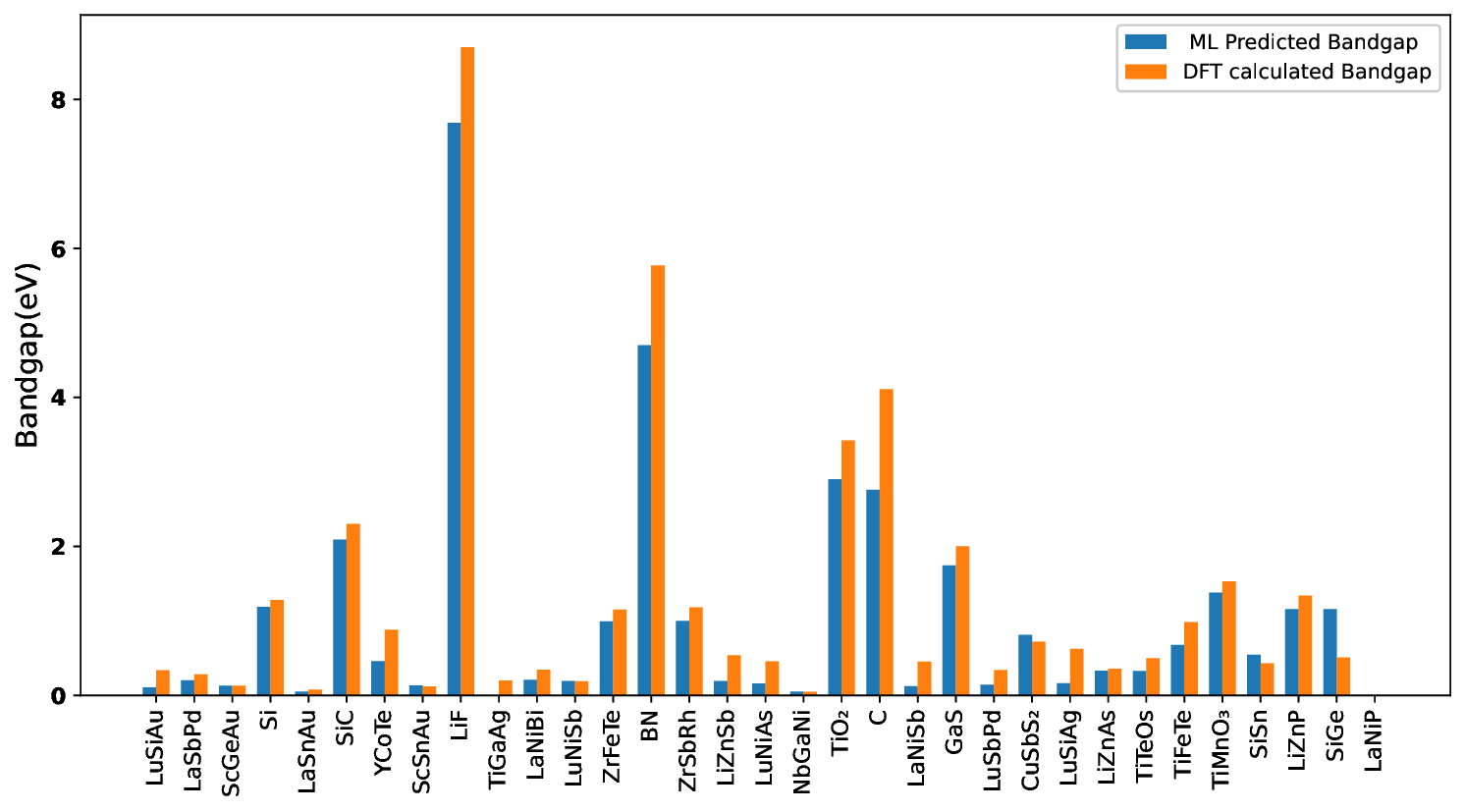}
          \label{fig:barplot}
     \end{subfigure}
     \begin{subfigure}[b]{0.2\textwidth}
         \centering
         \caption{ }
         \includegraphics[height=5cm]{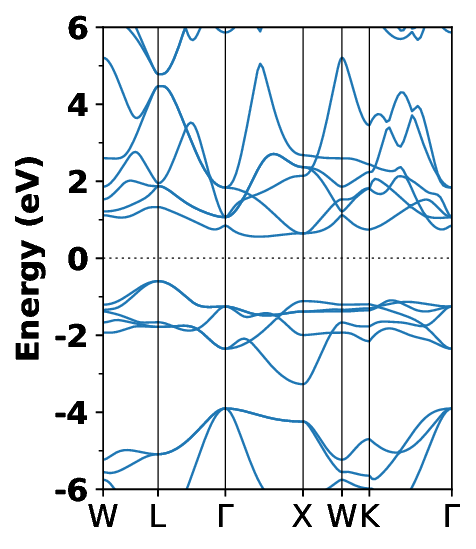}
         \label{fig:bandZr}
     \end{subfigure}
     \hfill
     \begin{subfigure}[b]{0.2\textwidth}
         \centering
         \caption{}
         \includegraphics[height=5cm]{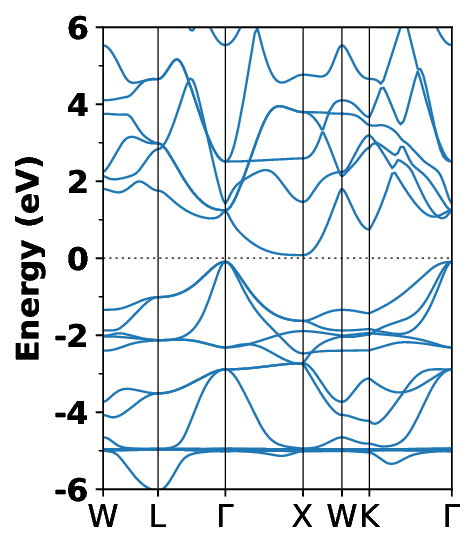} 
         \label{fig:bandLu}
     \end{subfigure}
     \hfill
     \begin{subfigure}[b]{0.2\textwidth}
         \centering
         \caption{ }
         \includegraphics[height=5cm]{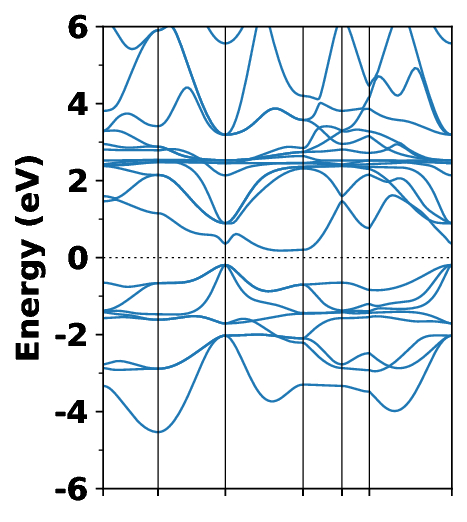}
         \label{fig:bandLa}
     \end{subfigure}
     \hfill
     \begin{subfigure}[b]{0.2\textwidth}
         \centering
         \caption{}
         \includegraphics[height=5cm]{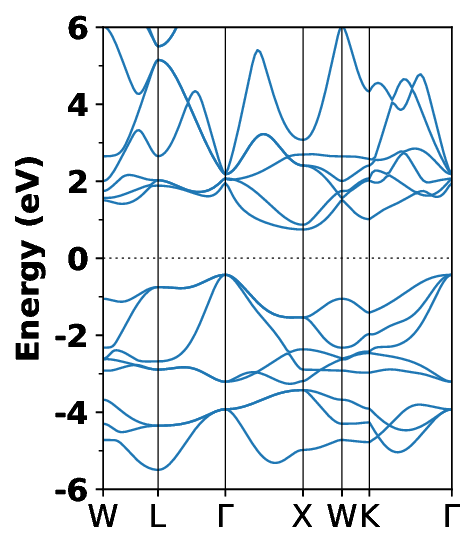}
         \label{fig:ZrSb}
     \end{subfigure}
     \caption{(a) The comparison between ML$-$predicted bandgap value obtained from best performing ML model i.e bagging with DT as base model and DFT$-$calculated bandgap values. The calculated electronic structure of newly predicted HH alloys (b) ZrFeTe, (c) LuNiSb (d) LaNiBi and (e) ZrSbRh. Obtained from GGA$-$PBE calculation.}
      \label{fig:band}
\end{figure*}

The R$^2$ values for the test and train datasets for the LassoCV as a base model in the bagging regression were just 0.511 and 0.509, respectively. These findings imply that the targeted property's variation is only moderately explained by the bagging regression with LassoCV as base model. Further, this model's predicting capability of the targeted property will be relatively poor, as seen by the higher RMSE values of 1.453/1.456 eV for test/train dataset, which represent large prediction errors.

In the bagging model with LassoCV as a base automatically sets coefficients of irrelevant features as zero, leading to loss of information. In contrast, RidgeCV retains all features but shrinks their coefficients simultaneously toward zero \cite{melkumova2017comparing}. The dropping of features in LassoCV can cause over simplification of the model and results in under fitting the data. If we use the LGBM as base model in our bagging regression, we found that, it gives reasonable performance, achieving R$^2$ values of 0.901/0.914 for the test/train datasets. This suggest that the LassoCV as a base model substantially capture the correlation between features and targeted property i.e, GGA$-$PBE bandgap value here. The corresponding RMSE values of 0.654/0.608 eV for test/train datasets suggest relatively low prediction error compared with that we obtain from LassoCV model. If we use the RidgeCV as base model in our bagging regression, we obtained modest performance with R$^2$ values of 0.676/0.678 for the test/train data-sets. These values suggest that RidgeCV as a base model has moderate ability to capture the correlation between features and the targeted property. Moreover, as the large value of RMSE  of 1.182/1.179 eV indicate larger prediction error which suggests that this model is unreliable to predict the targeted property. Finally, the XGB as base model performed well, yielding high R$^2$ values of 0.927/0.954 for the test/train datasets indicating good agreement between predicted and the actual values. The relatively low RMSE values of 0.561/0.444 eV for the test/train datasets further confirm the accuracy of the model to predict the GGA$-$PBE generated bandgap values.
\par
We have also explored the performance of various stacking models by combining different combinations of stand-alone ML regressor models considered in the present study. Among these models, one that stood out used RidgeCV as the meta-model, while employing LGBM, RF, and XGB as the base models. This particular stacking model exhibited promising results, with R$^2$ values of 0.943 and 0.991 for the test and train datasets, respectively. Additionally, this model demonstrated high accuracy, as evidenced by the low RMSE value of 0.494 eV for the test dataset and 0.199 for the train dataset. These RMSE values indicate that this model achieved remarkably small prediction errors, further emphasising its effectiveness.

Similarly, the stacking model consisting of RidgeCV, LGBM, RF as base models and LassoCV as meta$-$model demonstrated good performance, obtaining R$^2$ values of 0.942/0.993 for the test/train datasets. The corresponding RMSE values of 0.500/0.179 eV for the test/train dataset indicating rather small prediction errors,  demonstrating the efficacy of this stacking model. Additionally, the stacking model with SVR, DT, and KNN as base models and GB as the meta$-$model showed outstanding performance, achieving R$^2$ values of 0.899/0.981 for testing/training dataset. This stacking model's accuracy was further evidenced by the corresponding smaller RMSE values of 0.661/0.286 eV for the test/train datasets, which showed modest prediction errors. The stacking model where Linear regression, RF, GB are taken as base models and GB as meta$-$model gave us R$^2$ values of 0.943/0.944 for the test/train datasets, with RMSE values of 0.498/0.166 eV for the test/train datasets, showing modest prediction errors.These results suggest that the stacking models enhance prediction accuracy compared to stand-alone ML models, though to a slightly lesser extent than the bagging models mentioned above.

\begin{figure*}[t]
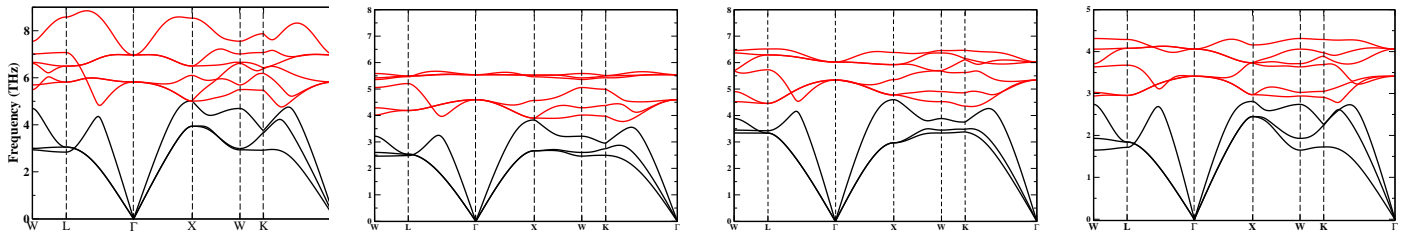

     \centering
     \begin{subfigure}[b]{0.21\textwidth}
         \centering
         \caption{ }
         \includegraphics[height=3cm]{PhonoZrFeTe.eps}
         \label{fig:phononZr}
     \end{subfigure}
     \hfill
     \begin{subfigure}[b]{0.21\textwidth}
         \centering
         \caption{}
         \includegraphics[height=3cm]{PhonoLuNiSb.eps} 
         \label{fig:phonLu}
     \end{subfigure}
     \hfill
     \begin{subfigure}[b]{0.21\textwidth}
         \centering
         \caption{ }
         \includegraphics[height=3cm]{PhonoZrSbRh.eps}
         \label{fig:phnZrSb}
     \end{subfigure}
     \hfill
          \begin{subfigure}[b]{0.21\textwidth}
         \centering
         \caption{ }
         \includegraphics[height=3cm]{PhonoLaNiBi.eps}
         \label{fig:phnLa}
     \end{subfigure}
     \caption{The calculated phonon dispersion curves of selected semiconducting compounds predicted by our ML model (a) ZrFeTe, (b) LuNiSb, (c) ZrSbRh, and (d) LaNiBi.}
     \label{fig:phn}
\end{figure*}
                                     
In Fig.\ref{fig:stac}, scatter plots are shown for above mentioned four different stacking models with averaged regression metrics and the predictions corresponding to the train/test dataset4 where the test and train inputs are shown by orange and blue circles, respectively. 

As a measure of comparison for the accuracy of different ML models considered in the present study, the targeted property from the best performing ML models are compared for new compounds those inputs are not considered to train the model. The best performing bagging and stacking model predicted bandgap values are validated using the DFT calculation using VASP code and are compared in Table~\ref{tab:stacking_models}. From this table it is clear that the bandgap values derived from GGA$-$PBE based DFT calculations are in good agreement with predictions of these models.

We would like to emphasise that, further improvements in the predicting capability of these ML models can be achieved by including additional features, new model selection, augment more data etc. Additionally, incorporating domain$-$specific knowledge and combining ML with physics$-$based models, such as DFT, could lead to more reliable and accurate predictions in the future. Upon a more detailed analysis of the results obtaned from ML predictions with those obtained from DFT calculations we have observed several patterns and trends. In general, both the Bagging and Stacking models exhibit a noticeable degree of deviation in predicting the bandgap value obtained from GGA$-$PBE based DFT calculations across various compositions. This indicates that there are inherent complexities and nuances in the dataset those are not adequately captured by the stand-alone ML models.

 Among the various ML models considered in the present study, best performing bagging model typically performs better than all stand-alone ML models and the stacking models in terms of predicting the bandgap values close to the computed GGA$-$PBE values for different compositions. For example, in the case of LuSiAu,  the bagging model predicts a bandgap value of 0.10 eV, Whereas, the stacking model predicts a value of 0.07 eV. It may be noted that the bandgap prediction from the bagging model is much closer to that from best performing stacking model. However these values are greatly depart from those obtained from GGA$-$PBE value of 0.337 eV in the present study. Moreover, it is important to note that, the deviations between the ML predicted bandgap value and that from GGA$-$PBE vary across different compositions. Some compositions exhibit relatively smaller deviations, indicating a better agreement between the ML models and the calculated bandgap values from DFT calculations as shown in Fig. \ref {fig:barplot}. Examples of such compositions include ScGeAu, Si, LiZnP, CuSbS$_2$, TiO$_2$, and SiC. Conversely, compositions such as  LuSbPd, LuSiAu, BN, C, and LiZnP demonstrate larger disparities between the bandgap value obtained from ML models and DFT calculations. This indicates that these compositions may possess unique or complex characteristics that are challenging for our trained ML models to capture accurately based on the features we have used to train the model or due to limitations in the dataset used to train our models.
\par

The observed deviations in ML predicted and GGA$-$PBE calculated values highlight the limitations of relying solely on ML techniques for predicting material properties. Even though ML models are powerful in many areas, they struggle with understanding complex quantum effects that greatly impact material properties. These effects are crucial for accurately getting material properties such as bandgap values considered in the present study which can be calculated with reasonable accuracy from DFT calculations. The deviation in the bandgap values between ML predicted and  DFT calculated highlight the need to be careful when looking at ML$-$generated predictions. While ML can quickly give predictions, it might not fully grasp the intricate quantum behaviors in complex materials. So, it's wise to be cautious when using ML predictions in situations where being really accurate is important.
In a nutshell, comparing the bandgap valued from ML predictions to DFT calculations reminds us that while ML  may provide insights, it cannot completely replace the quantum approaches to predict the material properties. Therefore, it is important to keep in mind that we have to employ ML predictions carefully and that they should complement, not replace the more precise quantum approaches. Given the many ways these approaches function to predict the properties of materials the ML approaches may be used to search wide chemical space to identify potential functional materials and followed by that use time consuming quantum approach such as DFT to predict material properties with reasonably accuracy. We believe that such hybrid approaches will be replaced with advanced ML models in future once developed more ML models and more reliable datasets to predict targeted properties.
\par
To improve the predictive accuracy of ML models for material properties, further research and development are necessary. This could involve exploring advanced feature engineering techniques, incorporating additional domain knowledge, and leveraging hybrid approaches that combine ML with DFT/empirical $-$based models. Further, if a  model is trained to predict the GGA$-$PBE bandgap values of materials, it is capable to predict GGA bandgap alone for new materials not for bandgap from experimentally or more computationally intensive DFT methods such as GW, hybrid etc. It may be noted that the results obtained from bagging and stacking models still exhibit notable discrepancies compared to calculated DFT values. Continued efforts to refine and enhance these models along with the integration of complementary approaches will be essential to achieve more reliable and accurate predictions in the future.

\section{Density Functional Theory Calculations}

The DFT calculations for structural optimization and the electronic structure calculations were performed using projector$-$augmented plane$-$wave (PAW)~\cite{kresse1999ultrasoft} method, as implemented in the Vienna \textit{ab$-$initio} simulation package (VASP)~\cite{kresse1996efficiency}. The generalized gradient approximation (GGA$-$PBE)~\cite{perdew1996generalized} proposed by Perdew$-$Burke\\$-$Ernzerh has been used for the exchange$-$correlation potential to compute the equilibrium structural parameters and electronic structure. The irreducible part of the first  Brillouin zone (IBZ) was sampled using a Monkhorst pack scheme~\cite{monkhorst1976special} and employed a 12$\times$12$\times$12  \textbf{k}$-$mesh for geometry optimization. 
\par
A plane$-$wave energy cutoff of 600 eV is used for geometry optimization. The convergence criterion for energy was taken to be 10$^{-6}$ eV/cell for electronic minimisation step  and less than 1 meV/\AA~ is used for Hellmann$-$Feynman force acting on each atom to find the equilibrium atomic positions. Our previous study~\cite{choudhary2020thermal} shows that the computational parameters used for the present study are sufficient enough to accurately predict the equilibrium structural parameters for HH compounds. In our electronic structure calculations we focused on the high symmetry directions of the first IBZ of face centered cubic systems. Considering that the GGA$-$PBE method tends to underestimate bandgap values, we have also included results from higher level calculations, such as the TB$-$mBJ method in tablexx though it cannot be compared with ML prediction values. The other theoretical details of these calculations can be found elsewhere~\cite{choudhary2023first,tran2009accurate,koller2011merits,koller2012improving}. 
In order to see the dynamical stability of the predicted four HH alloys based on 18 VEC, we have made lattice dynamic calculation to check for their dynamic stability. The  phonon dispersion curves are generated using the  finite displacement method implemented in the VASP$-$phonopy~\cite{togo2015first} interface with the supercell approach. We have used the  relaxed primitive cells to create supercells of dimension 3 × 3 × 3 with the displacement distance of 0.01 Å for our finite displacement calculations to estimate the dynamical matrix. 

In Table \ref{tab:stacking_models} we have listed out the predicted bandgap values from our ML regression models discussed in section \ref{Machine Learning Algorithm} along with those obtained from DFT calculation mentioned above. It may be noted that, the HH compounds listed in Table \ref{tab:stacking_models} are predicted based on 18 VEC rule and their corresponding bandgap values obtained from GGA$-$PBE calculations are directly compared with those predicted from our best performing ML models. This comparison is made to validate our ML trained models performance and also their reliability to predict GGA bandgap value of any unknown new compounds. Additionally, we have predicted the GGA$-$PBE bandgap values for well$-$studied compounds such as Si, C, SiC, LiF, BN, TiO$_2$, GaS, SiGe and SiSn (see Fig. \ref {fig:barplot}) using the best performing ML models to show their reliability to predict GGA bandgap value of wide range of semiconductors. The overall observation is that,  we have obtained  minimal error between the bandgap values calculated using the GGA$-$PBE method and those predicted based on our trained ensemble ML models for well known semiconductors.  
\par
 The calculated electronic band structures of newly identified HH compounds ZrFeTe, LuNiSb, ZrSbRh and LaNiBi are displayed in Fig.\ref{fig:band} as (a),(b), (c), (d), respectively. From these figures it is clear that all these compounds possess indirect bandgap behavior with the valence band maximum (VBM) located at the $\Gamma$ point and the conduction band minimum (CBM) at the X point. At the $\Gamma$ point, the bands near the VBM show a triply degenerate state. The degenerate bands near the VBM create a well dispersed band structure that favors hole conductivity in these systems. However, the flat band dispersion near the CBM along the $\Gamma$ $-$X direction suggests low electron conductivity due to the high effective mass of electrons. 
 \par

In order to check the phase stability of our selected compounds, we have  made enthalpy of formation ($\Delta H_f$) calculations. Our calculated 
$\Delta H_f$ is negative for ZrFeTe, LuNiSb, ZrSbRh and LaNiBi with the value of -0.76 eV/atom, -0.88 eV/atom, -1.0 eV/atom and -0.77 eV/atom, respectively. The negative $\Delta H_f$ values confirm that these materials are stable at ambient conditions and also can be synthesised experimentally.  Additionally, to check the dynamical stability of those compounds we have calculated the phonon dispersion curves along high symmetry directions within the fcc BZ.
We used the Phonopy code with the GGA$-$PBE functional and the finite difference method to compute the phonon dispersion at equilibrium lattice parameters under harmonic approximation. The phonon dispersion curve provides valuable information about the dynamical stability of a crystal. A dynamically stable crystal exhibits only positive (real) phonon frequencies throughout the phonon dispersion curve. In contrast, a dynamically unstable crystal also display negative (imaginary) phonon frequencies. The calculated phonon dispersion curves for predicted new HH alloys are given in Fig.\ref{fig:phn}. From these phonon dispersion curves we found that no one of these compounds exhibit a dynamical instability and hence no negative frequency is found in any of these compounds. This result suggests that all these compounds will be stable under the ambient condition.
 The observation of high stability in these newly identified HH alloys is consistent with the alloy theory of solids that 18 VEC rule brings extra stability to the system. 

\section{Conclusions}

In summary, our trained best performing machine learning models exhibited minimal deviations from actual values when predicting GGA$-$PBE bandgap values using structural and composition based features. When compared to DFT calculations, ML models trained on existing DFT data can provide comparable prediction accuracy in significantly less time. These ML models can effectively aid in the search for compositions with desirable bandgap value across a wide chemical space, which accelerate material discovery with desired properties. Among the 8 stand-alone ML regression models we have studied, Adaboost performed the poorest. Conversely, the DT model showed signs of over-fitting on the training data, as indicated by its performance matrices. The other stand-alone models consistently improved their performance when we increase the  dataset size, emphasising their robustness, accuracy, and generalization capabilities. Notably, CatBoost regression excelled on larger datasets, while LGBM regression demonstrated strong performance on smaller datasets. RF regression consistently delivered reliable results on smaller and medium$-$sized datasets. GB performed well across datasets of varying sizes, while XGB exhibited mixed performance depending on the dataset size. It is important to note that these results depend on the specific dataset characteristics, feature engineering techniques, and pre$-$processing steps used. More exploration and experimentation with different combinations of features and pre$-$processing techniques can further enhance the predictive capabilities of these models. 

The selection of an appropriate model should be based on a comprehensive understanding of the dataset and the specific requirements of the prediction task. Subsequently, we employed above mentioned stand alone ML models to predict the bandgap values of new HH compounds designed based on 18 VEC. Through this process, we identified several new materials with significant potential for applications in thermoelectrics, solar cells, and photocatalysis. In addition, we employed ensemble model such as stacking and bagging models to enhance the predicting capability for obtaining the GGA$-$PBE bandgap values for unseen datasets. It's important to take into account that in this study, we adopted  GGA$-$PBE bandgap values to train our ML models. For attaining a superior level of accuracy to predict the bandgap value on par with that measured experimentally,  we have to train our ML models with experimental bandgap values which is available only for very limited number of systems. Also, the best performing ML models can be  trained using the bandgap values obtained from more advanced DFT calculations such as hybrid functionals, GW method etc. However the bandgap values obtained from such advanced DFT methods are very limited owing to the fact that such calculation are computationally very intensive. Moreover, the calculated phonon dispersion curve and negative $\Delta H_f$ in all the  selected compounds considered in this study confirmed their phase and dynamic stability at ambient condition. We hope that the present approach hold the potential to greatly expedite the discovery of new materials possessing desirable properties for various energy$-$related applications.

\section{Conflicts of interest}
There are no conflicts to declare.
\section*{Acknowledgements}
The authors are grateful to the SCANMAT Centre, Central University of Tamil Nadu, Thiruvarur for providing the computer time at the SCANMAT supercomputing facility. The authors would also like to acknowledge the SERB$-$Core Research Grant (CRG) vide file no.CRG/2020/001399.

\bibliography{3paperbib}
\end{document}